\def\fr#1#2{\hbox{${#1\over #2}$}}
\def\subsub#1{\medskip{\bf #1}}
\def\lra{\leftrightarrow}            \def\Ra{\Rightarrow}
\def\ni{\noindent}                   \def\pd{\partial}
\def\ort{\perp}                      \def\wtilde{\widetilde}
\def\what#1{\hat #1{}}               
         \def\tgr{GR$_{\parallel}$}
\def\mb#1{\hbox{\boldmath$#1$}}      \def\bG{\mb{\Gamma}}
\def\bT{\bar T}                      \def\bcL{{\bar{\cal L}}}
\def\bcH{{\bar{\cal H}}}             \def\bphi{\bar\phi}
\def\bchi{\bar\chi}                  

\def\m{\mu}             \def\n{\nu}              \def\k{\kappa}
\def\G{\Gamma}          \def\g{\gamma}           \def\d{\delta}
\def\S{\Sigma}          \def\s{\sigma}           \def\t{\tau}
\def\a{\alpha}          \def\b{\beta}            \def\th{\theta}
\def\vphi{\varphi}      \def\ve{\varepsilon}
\def\r{\rho}            \def\D{\Delta}           \def\p{\pi}
\def\l{\lambda}         \def\o{\omega}           \def\O{\Omega}
\def\bn{\bar{n}}        \def\bi{{\bar i}}        \def\bk{{\bar k}}
       \def\bm{{\bar m}}        \def\bn{{\bar n}}
\def\bj{{\bar j}}
\def\cL{{\cal L}}       \def\cH{{\cal H}}        \def\cP{{\cal P}}
\def\cJ{{\cal J}}       \def\cO{{\cal O}}        \def\bu{\bar u}
\def\hp{{\hat\pi}}      \def\tcL{\tilde\cL}
\documentclass[12pt]{article}
\usepackage{latexsym,graphicx,equations}
\def\nn{\nonumber}
\def\be{\begin{equation}}             \def\ee{\end{equation}}
\def\bea{\begin{eqnarray} }           \def\eea{\end{eqnarray} }
\def\lab#1{\label{eq:#1}}             \def\eq#1{(\ref{eq:#1})}
\def\bsubeq{\begin{subequations}}     \def\esubeq{\end{subequations}}
\def\bitem{\begin{itemize}}           \def\eitem{\end{itemize}}
\newcommand{\sect}[1]{\addtocounter{section}{1}
      \section*{\large\thesection.~#1}\setcounter{equation}{0}}
\newcommand{\ack}[1]{\section*{\large #1}}

\def\bull{\rule[.4ex]{1ex}{.45ex}}    %\hskip1.1ex}
\begin{document}
\title{\vskip-1cm\large\bf Hamiltonian structure and gauge symmetries\\
                 of Poincar\'e gauge theory\thanks{
    Invited talk presented at the Meeting of German Physical Society,
    Dresden, March 20--24, 2000.} }
\author{\normalsize M. Blagojevi\'c\thanks{
                    Email address: mb@phy.bg.ac.yu}\\
        \small {\it Institute of Physics, 11001 Belgrade,
                    P. O. Box 57, Yugoslavia}}
\date{}
\maketitle
\begin{abstract}
This is a review of the constrained dynamical structure of
Poincar\'e gauge theory which concentrates on the basic canonical and
gauge properties of the theory, in\-clu\-ding the identification of
constraints, gauge symmetries and conservation laws. As an interesting
example of the general approach we discuss the teleparallel formulation
of general relativity.
\end{abstract}

\sect{Introduction}

Despite its successes in describing known macroscopic gravitational
phenomena, Einstein's general relativity (GR) still lacks the status of
a fundamental microscopic theory. This is so because the theory admits
singular solutions under very general assumptions, and all attempts to
quantize GR encounter serious difficulties. Among various attempts to
overcome these problems, gauge theories of gravity are especially
attractive, as they are based on the concept of gauge symmetry which
has been very successful in the foundation of other fundamental
interactions. The importance of the Poincar\'e symmetry in particle
physics leads one to consider the Poincar\'e gauge theory (PGT) as a
natural framework for a description of the gravitational phenomena
[1--6] (for more general attempts, see \cite{7}).

In this paper we give an introduction to PGT and general Dirac's
formalism, and a review of the constrained dynamical structure of PGT,
aimed at a fairly general audience. We begin our exposition by giving a
short outline of PGT in Sec. 2. In the next section we introduce basic
concepts of Dirac's Hamiltonian formalism, which are essential for a
clear understanding of the canonical and gauge structure of any gauge
theory. Then, we discuss the construction of the Hamiltonian and the
algebra of constraints in the general PGT (Sec. 4), derive the related
generator of Poincar\'e gauge symmetry (Sec. 5), and introduce the
important concept of energy (and other conserved charges) on the basis
of the asymptotic structure of spacetime (Sec. 6). The teleparallel
form of GR is an interesting limit of PGT, both theoretically and
observationally; its canonical properties are discussed in Sec. 7.
Section 8 is devoted to concluding remarks.

Our general conventions, with some exceptions in Sect. 3, are the
following: the Latin indices refer to the local Lorentz frame, whereas
the Greek indices refer to the coordinate frame; the first letters of
both alphabets $(a,b,c,\dots;\a,\b,\g,\dots)$ run over $1,2,3$, while
the middle alphabet letters $(i,j,k,\dots;\m,\n,\l,\dots)$ run over
$0,1,2,3$; $\eta_{ij}=(+,-,-,-)$, and $\ve^{ijkl}$ is completely
antisymmetric tensor normalized to $\ve^{0123}=+1$.

\sect{Poincar\'e gauge theory}

It is well known that the existence and interaction of certain fields,
such as the electromagnetic field, can be closely related to invariance
properties of the theory. Thus, for instance, if the Lagrangian of
matter fields is invariant under phase transformations with
{\it constant\/} parameters $\a$, the electromagnetic field can be
introduced by demanding the invariance under the extended,
{\it local\/} transformations, obtained by replacing $\a$ with a function
of spacetime points, $\a\to\a(x)$.

On the other hand, it is much less known that Einstein's GR is
invariant under local Poincar\'e transformations. This property is
based on the principle of equivalence, and gives a rich physical
content to the concept of local (or gauge) symmetry. Instead of
thinking of local Poincar\'e symmetry as derived from the principle of
equivalence, the whole idea can be reversed, in accordance with the
usual philosophy of gauge theories. When gravitational field is absent,
it has become clear from a host of experiments that the underlying
symmetry of fundamental interactions is given by the Poincar\'e group.
If we now want to make a physical theory invariant under {\it local\/}
Poincar\'e transformations, it is necessary to introduce new, {\it
compensating\/} (or gauge) fields, which, in fact, represent
the gravitational interaction [1--6].

\subsub{Global Poincar\'e symmetry.} In physical processes at low
energies gravitational field does not have a significant role, since
the gravitational interaction is extremely weak. The spacetime without
gravity is described by the special relativity theory (SR), and its
mathematical structure corresponds to Minkowski space $M_4$. Any
physical observer in $M_4$ uses some {\it reference frame\/}, endowed
with coordinates $x^\m$ $(\m=0,1,2,3)$, serving to identify
physical events. An {\it inertial\/} observer can always choose global
inertial coordinates, such that the infinitesimal interval has the form
$ds^2=\eta_{\m\n}dx^\m dx^\n$, where $\eta_{\m\n}=(1,-1,-1,-1)$ is
the metric tensor. The equivalence of inertial reference frames is
expressed by the global (rigid) Poincar\'e symmetry $P(1,3)$ in $M_4$.

At each point of $M_4$, labeled by coordinates $x^\m$, one can define
an inertial laboratory by a local Lorentz frame --- a set of four
orthonormal, tangent vectors $\mb{e}_i(x)$,
$\mb{e}_i\cdot\mb{e}_j=\eta_{ij}$, called the tetrad. In global
inertial coordinates $x^\m$, one can always choose the tetrad such that
it coincides with a coordinate frame $\mb{e}_\m(x)$ (a set of four
vectors, tangent to coordinate lines at $x$), i.e.
$\mb{e}_i=\d_i^\m\mb{e}_\m$. Here, the Latin indices $(i,j,...)$ refer
to local Lorentz frames, while the Greek indices $(\m,\n,...)$ refer to
coordinate frames.

The physics of fundamental interactions is successfully described by
Lagrangian field theory. Dynamical variables in this theory are fields
$\phi(x)$, and dynamics is determined by a function of fields and their
derivatives, $\cL(\phi,\pd_k\phi)$, called the Lagrangian. The action
integral $I=\int d^4 x\cL$ is invariant under an arbitrary spacetime
transformation $x'^m=x^m+\xi^m$, $\phi'(x)=\phi(x)+\d_0\phi(x)$,
provided the Lagrangian satisfies the condition \cite{1}
\be
\d_0\cL+\pd_m(\xi^m\cL)=0\, ,                               \lab{2.1}
\ee
where $\d_0\cL\equiv(\pd\cL/\pd\phi)\d_0\phi
              +(\pd\cL/\pd\phi_{,k})\d_0\phi_{,k}$.
Note that this condition can be, in fact, relaxed, by allowing an
arbitrary four--divergence to appear on the right hand side. The
Lagrangian $\cL$ satisfying the above invariance condition is called
an invariant density.

Consider now a matter field $\psi(x)$ in spacetime, referred
to a local Lorentz frame. Its transformation law under the global
Poincar\'e transformations $\xi^m=\o^m{_n}x^n+\ve^m$ has the
form
$$
\d_0\psi=\bigl(\fr{1}{2}\o\cdot M+\ve\cdot P\bigr)\psi
        =\bigl(\fr{1}{2}\o\cdot\S+\xi\cdot P\bigr)\psi\, ,
$$
where $M_{mn}=x_m\pd_n-x_n\pd_m+\S_{mn}$ and $P_m=-\pd_m$ are the
Poincar\'e generators, and $\S_{mn}$ is the spin matrix. Using the
invariance condition (2.1), one finds that the invariance of the
Lagrangian $\cL_M(\psi,\pd_k\psi)$ under $P(1,3)$ leads to the
conservation of energy--momentum and angular momentum currents.

Localization of Poincar\'e symmetry leads to Poincar\'e gauge theory of
gravity, in which {\it both\/} energy--momentum and spin currents of
matter fields are naturally included into gravitational dynamics, in
contrast to GR.

\subsub{Local Poincar\'e symmetry.} We now describe the process
of transition from global to local Poincar\'e symmetry, and find its
relation to the gravitational interaction.  Other spacetime symmetries
[de Sitter, conformal, etc.] can be treated in an analogous manner.

Consider a Lagrangian for matter fields, $\cL_M=\cL_M(\psi,\pd_k\psi)$,
defined with respect to a local Lorentz frame and invariant under
$P(1,3)$. If we now generalize global Poincar\'e transformations by
replacing ten {\it constant\/} group parameters $\o^m{_n},\ve^m$ with
some {\it functions\/} of spacetime points, the invariance condition
\eq{2.1} is violated. The violation of (local) invariance can be
compensated by certain modifications of the original theory. After
introducing the {\it covariant derivative\/},
\bsubeq\lab{2.2}
\be
\nabla_k\psi=h_k{^\m}\nabla_\m\psi = h_k{^\m}\bigl(\pd_\m
             +\fr{1}{2}A^{ij}{_\m}\S_{ij}\bigr)\psi\, ,     \lab{2.2a}
\ee
where $h_k{^\m}$ and $A^{ij}{_\m}$ are the compensating fields,
the modified, invariant Lagrangian for matter fields takes the form
\be
\wtilde\cL_M =b\cL_M(\psi,\nabla_k\psi) \, ,                \lab{2.2b}
\ee
\esubeq
where $b=\det(b^k{_\m})$, and $b^k{_\m}$ is the inverse of $h_k{^\m}$.
It is obtained from the original Lagrangian $\cL_M(\psi,\pd_k\psi)$ in
two steps:
\bitem
\item[\bull] by replacing $\pd_k\psi\to \nabla_k\psi$
     (minimal coupling), and \par
\item[\bull] multiplying $\cL_M$ by $b$.
\eitem
\ni The Lagrangian $\wtilde\cL_M$ satisfies Eq. \eq{2.1} by construction,
hence it is an invariant density.

The matter Lagrangian is made invariant under local transformations by
introducing compensating fields. In order to construct an invariant
Lagrangian for the new fields $b^k{_\m}$ and $A^{ij}{_\m}$, we
introduce the corresponding {\it field strengths\/},
\bea
&&F^{ij}{_{\m\n}}=
  \pd_\m A^{ij}{_\n}+A^i{_{s\m}}A^{sj}{_\n}-(\m\lra\n)\, ,\nn\\
&&F^i{_{\m\n}}=
  \pd_\m b^i{_\n}+A^i{}_{s\m}b^s{_\n}-(\m\lra\n)\, ,        \lab{2.3}
\eea
related to the Lorentz and translation subgroups of $P(1,3)$,
respectively. The new Lagrangian must be an invariant density
depending only on the Lorentz and translation field strengths, so that
the complete Lagrangian of matter and gauge fields has the form
\be
\wtilde\cL=b\cL_F(F^{ij}{_{kl,}}F^i{_{kl}})+b\cL_M(\psi,\nabla_k\psi)
                                                      \, . \lab{2.4}
\ee

Up to this point, we have not given any geometric interpretation to the
new, compensating fields $b^k{_\m}$ and $A^{ij}{_{\m}}$. Such an
interpretation is possible and useful, and it leads to a new
understanding of gravity.

\subsub{Riemann--Cartan geometry.} Now, we introduce some basic
geometric concepts in order be able to understand the geometric
meaning of PGT \cite{2,8}.

Spacetime is often described as a ``four--dimensional continuum". In
SR, it has the structure of Minkowski space $M_4$. In the presence of
gravity spacetime can be divided into ``small, flat pieces"  in which
SR holds (on the basis of the principle of equivalence), and these
pieces are ``sewn together" smoothly. Although spacetime looks locally
like $M_4$, it may have quite different global properties. Mathematical
description of such four--dimensional continuum is given by the concept
of a {\it differentiable manifold\/}.

We assume that spacetime has a structure of a differentiable manifold
$X_4$.  On $X_4$ one can define differentiable mappings, tensors, and
various algebraic operations with tensors at a given point (addition,
multiplication, contraction). However, comparing tensors at different
points requires some additional structure on $X_4$: the law of parallel
transport, defined by a linear (or affine) {\it connection\/} \bG. An
$X_4$ equipped with \bG\ is called {\it linearly connected space\/},
$L_4=(X_4,\bG)$. Linear connection is equivalently defined by the
covariant derivative \mb{D}; thus, for instance,
$D_\r A^\m=\pd_\r A^\m +\G^\m_{\l\r}A^\l$.

Linear connection is not a tensor, but its antisymmetric part defines a
tensor called the torsion tensor:
$T^{\m}{_{\l\r}}=\G^\m_{\r\l}-\G^\m_{\l\r}$.

Parallel transport is a path dependent concept. If we parallel
transport a vector around an infinitesimal closed path, the result is
proportional to the Riemann curvature tensor: $R^{\m}{_{\n\l\r}}
=\pd_\l\G_{\n\r}^\m +\G_{\s\l}^\m\G_{\n\r}^\s-(\l\lra\r)$.

On $X_4$ one can define {\it metric tensor} \mb{g} as a symmetric,
nondegenerate tensor field of type $(0,2)$. After that we can
introduce the scalar product of two tangent vectors and calculate
lengths of curves, angles between vectors, etc. Linear connection and
metric are  geometric objects independent of each other. The
differentiable manifold $X_4$, equipped with linear connection and
metric, becomes {\it linearly connected metric space\/} $(L_4,\mb{g})$.

In order to preserve lengths and angles under parallel transport in
$(L_4,\mb{g})$, one can impose the {\it metricity condition\/}
\be
-Q_{\m\n\l}\equiv D_{\m}g_{\n\l}=\pd_{\m}g_{\n\l}
       -\G_{\n\m}^{\r}g_{\r\l}-\G_{\l\m}^{\r}g_{\n\r}=0\, , \lab{2.5}
\ee
which relates $\bG$ and \mb{g}. The requirement of vanishing
nonmetricity \mb{Q} establishes local Minkowskian structure
on $X_4$, and defines a metric compatible linear connection.

A space $(L_4,\mb{g})$ with the most general metric compatible linear
connection \bG\ is called {\it Riemann--Cartan\/} space $U_4$. If the
torsion vanishes, $U_4$ becomes {\it Riemann\/} space $V_4$ of GR; if,
alternatively, the curvature vanishes, $U_4$ becomes Weitzenb\"ock's
{\it teleparallel\/} space $T_4$. Finally, the condition
$\,R^\m{_{\n\l\r}}=0\,$ transforms $V_4$ into Minkowski space $M_4$,
and $\,T^\m{_{\l\r}}=0\,$ transforms $T_4$ into $M_4$ (Figure 1).
\begin{figure}
\begin{center}
\includegraphics[height=4.5cm]{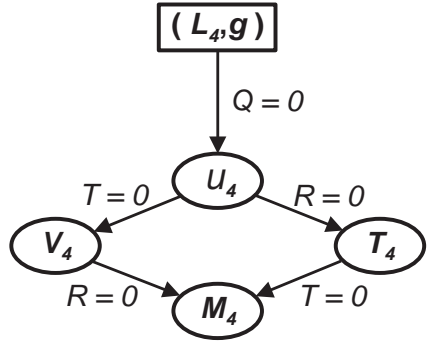}
\end{center}
\vspace*{-.5cm}
\caption{Classification of spaces satisfying the metricity condition}
\end{figure}

The choice of basis in a tangent space is not unique. Coordinate frame
$\mb{e}_\m$ and local Lorentz frame $\mb{e}_i$ are of particular
practical importance. Every tangent vector \mb{u} can be expressed in
both frames: $\mb{u}=u^\m\mb{e}_\m=u^i\mb{e}_i$. In particular,
$\mb{e}_i=e_i{^\m}\mb{e}_\m$ and $\mb{e}_\m=e^i{_\m}\mb{e}_i$. As a
consequence, we find that $u^i=e^i{_\m}u^\m$, $u^\m=e_i{^\m}u^i$. We
denote the connection components with respect to coordinate and local
Lorentz frames as $\G^\m_{\l\r}$ and $\o^{ij}{_\m}$, respectively.

Linear connection and metric are geometric objects independent of the
choice of frame. Their components are defined with respect to
a frame and are, clearly, frame--dependent.

\subsub{Interpretation of PGT.} The final result of the analysis
of PGT is the construction of the invariant Lagrangian \eq{2.4}. It is
achieved by introducing new fields $h_i{^\m}$ (or $b^k{_\n}$) and
$A^{ij}{_\m}$, which are used to construct the covariant derivative
$\nabla_k$ and the field strengths $F^{ij}{_{\m\n}}$ and
$F^i{_{\m\n}}$. This theory can be thought of as a field theory in
$M_4$. However, geometric analogies are so strong, that it would be
unnatural to ignore them. The following properties are essential for
the geometric interpretation of PGT:
\bitem
\item[\bull] the Lorentz gauge field $A^{ij}{_\m}$ can be identified
    with the Lorentz connection $\o^{ij}{_\m}$, or, equivalently, the
    covariant derivative $\nabla_\m(A)$ can be identified with
    $D_\m(\o)$;
\item[\bull] the translation gauge field $b^k{_\m}$ can be identified
     with the tetrad $e^k{_\m}$; \par
\item[\bull] local Lorentz symmetry of PGT implies the metricity
     condition \eq{2.5}.
\eitem
\ni Consequently, PGT has the geometric structure of Riemann--Cartan
space $U_4$.

It is not difficult to prove that the field strengths $F^i{_{\m\n}}$
and $F^{ij}{}_{\m\n}$ represent geometrically the torsion
$T^{\l}{_{\m\n}}$ and the curvature $R^{\l}{}_{\t\m\n}$, respectively.

Since Riemann--Cartan space has locally the structure of $M_4$, it
follows that PGT is consistent with the principle of equivalence
\cite{9}. Thus, PGT is a gauge approach to the theory of gravity,
possessing a transparent {\it geometric and physical\/} interpretation.

\sect{Constrained Hamiltonian dynamics}

Despite many successes of quantum theory in describing basic physical
phenomena, one is continually running into difficulties in some
specific physical situations. Thus, all attempts to quantize the theory
of gravity encounter serious difficulties. In order to find a solution
of these problems, it seems to be useful to reconsider the fundamental
principles of classical dynamics.  In this context, the principles of
Hamiltonian dynamics are seen to be of great importance for a
basic understanding of both classical and quantum theory.

In gauge theories the number of dynamical variables in the action is
larger than the number of physical variables.  The presence of
unphysical variables is closely related to the existence of gauge
symmetries. In the Hamiltonian formalism dynamical systems of this type
are characterized by the presence of constraints. Here, we present the
basic ideas of the constrained Hamiltonian dynamics \cite{10,11,12}.

\subsub{Primary constraints.} In order to simplify the exposition we
start by considering a classical system of particles, described
by coordinates $q_i\,\, (i=1,2,...,N)$ and the action
$I=\int dt\,L(q,\dot q)$. To go over to the Hamiltonian formalism we
introduce, in the usual way, the momentum variables:
\be
p_i={\pd L\over\pd\dot q_i}\equiv f_i(q,\dot q)
                               \qquad (i=1,2,...,N)\, .       \lab{3.1}
\ee
In simple dynamical theories these relations can be inverted, so that
{\it all\/} velocities can be expressed in terms of the coordinates
and momenta, whereupon one can simply define the Hamiltonian function.
However, for many interesting theories such an assumption would be too
restrictive. Therefore, we shall allow for the possibility that momentum
variables are not independent functions of velocities, i.e. that there
exist {\it constraints\/}:
\bsubeq\lab{3.2}
\be
\phi_m(q,p)=0 \qquad (m=1,2,...,P)\, .                      \lab{3.2a}
\ee
The variables $(q,p)$ are local coordinates on the phase space
$\G$. The relations (3.2a) are called {\it primary constraints\/}; they
determine a subspace $\G_1$ of $\G$, in which the time development of a
dynamical system takes place.

The geometric structure of the subspace $\G_1$ can be very complicated.
If $(i)$ the rank of the matrix $\cJ=\pd\phi_m/\pd(q,p)$ is constant on
$\G_1$, and $(ii)$ all the constraint functions $\phi_m$ are
independent (irreducible), then the Jacobian $\cJ$ is of rank $P$ on
$\G_1$. Accordingly, the dimension of $\G_1$ is well defined and equal
to $2N-P$.

\subsub{Weak and strong equalities.} At this stage we introduce the
useful notions of weak and strong equality. Let $F(q,p)$ be a function
which is defined and differentiable in a neighborhood
${\cO}_1\subseteq\G$ containing the subspace $\G_1$. If the restriction
of $F(q,p)$ on $\G_1$ vanishes, we say that $F$ is weakly equal to zero:
$F(q,p)\approx 0$. If the function $F$ and all its first derivatives
vanish on $\G_1$, than $F$ is strongly equal to zero: $F(q,p)=0$. For
strong equality we shall use the usual equality sign. This definition
is especially useful in the analysis of the equations of motion, which
contain derivatives of functions on $\G_1$.

By using these conventions the relations \eq{3.2a} can be written as
weak equalities:
\be
\phi_m(q,p)\approx 0      \qquad (m=1,2,...,P)\, .          \lab{3.2b}
\ee
\esubeq

It is now interesting to clarify the relation between strong and weak
equalities: if $F(q,p)$ vanishes weakly, $F\approx 0$, what can one say
about the derivatives of $F$ on $\G_1$? By use of the general method of
calculus of variations with constraints, one can show that $F\approx 0$
implies $F-\l^m\phi_m \approx \cO$, where $\l^m$ are some
multipliers, and $\cO$ is a quantity whose derivatives are weakly
vanishing: it can be zero, a constant, or second or higher power of a
constraint. For theories in which the constraint functions $\phi_m$
satisfy the above regularity condition $(i)$ and $(ii)$, one can show
that $\cO=0$ \cite{11}.

\subsub{Hamiltonian.} Consider now the quantity
\be
H_c=p_i\dot q_i-L(q,\dot q)\, .                              \lab{3.3}
\ee
By making variations in $q$ and $\dot q$ one finds that $H_c$ can be
expressed in terms of $q$'s and $p$'s only, independent of velocities.
Expressed in this way it becomes the {\it canonical\/} Hamiltonian.
Since $H_c$ is not uniquely defined in the presence of constraints,
it is natural to introduce the {\it total\/} Hamiltonian
\be
H_T=H_c+u^m\phi_m\, ,                                       \lab{3.4}
\ee
where $u^m$'s are, at this stage, arbitrary multipliers. By varying
this expression with respect to $(u,q,p)$ one obtains the constraints
\eq{3.2} and the Hamiltonian equations of motions involving arbitrary
multipliers. The Hamiltonian equations for an arbitrary dynamical
quantity $g(q,p)$ can be written in the form
\be
\dot g=\{ g,H_c\} +u^m\{ g,\phi_m\}\approx\{ g,H_T\}\, ,   \lab{3.5}
\ee
where $\{X,Y\}$ is the Poisson bracket (PB) of $X$ and $Y$.

Equations \eq{3.5} describe the motion of a system in the subspace
$\G_1$ of dimension $2N-P$. The motion is described by $2N$
coordinates $(q,p)$ satisfying $P$ constraints; as a consequence,
there appear $P$ multipliers in the evolution equations.
Explicit elimination of some coordinates is possible, but it may
lead to a violation of locality and/or covariance.

\subsub{Consistency conditions.} A basic consistency of the theory
requires that the primary constraints be conserved during the temporal
evolution of the system:
\be
\dot\phi_m =\{\phi_m,H_c\} +u^n\{\phi_m,\phi_n\}\approx 0\, .\lab{3.6}
\ee
If the equations of motion are consistent, the above conditions reduce
to the one of the following three cases:
{$a)$} an identity, 0=0;
{$b)$} an equation independent of the multipliers, yielding a
new, {\it secondary constraint\/}: $\chi(q,p)\approx 0$;
{$c)$} a restriction on the $u^n$'s.

If we find some secondary constraints in the theory, they also have to
satisfy consistency conditions of the type \eq{3.6}. The process
continues until all consistency conditions are exhausted. As a final
result we are left with a number of new constraints, and a number
of conditions on the multipliers.

Let us denote all the constraints in the theory as
$\vphi_s\equiv (\phi_m,\chi_n)\approx 0$. These constraints define a
subspace $\G_2$ of the phase space $\G$, such that $\G_2\subseteq
\G_1$. The notions of weak and strong equalities are now defined with
respect to $\G_2$.

Since the consistency requirements lead, in general, to some conditions
on the multipliers, the total Hamiltonian can be written in the form
\be
H_T= H' + v^a\phi_a \qquad (a=1,2,...,N'_1)                \lab{3.7}
\ee
where $H'=H_c+U^m\phi_m$, $U^m$ are determined and $v^a$ arbitrary
multipliers.

Thus, even after all the consistency requirements are satisfied, we
still have {\it arbitrary\/} functions of time in the theory. As a
consequence, dynamical variables at some future instant of time are
{\it not uniquely\/} determined by their initial values.

\subsub{First class and second class quantities.} A dynamical variable
$R(q,p)$ is said to be first class (FC) if it has weakly vanishing PBs
with all constraints in the theory: $\{ R,\vphi_s\}\approx 0$. If $R$
is not first class, it is called second class. While the distinction
between primary and secondary constraints is of little importance in
the final form of the Hamiltonian theory, the property of being FC or
second class is essential for the dynamical interpretation of
constraints.

We have seen that (in a regular theory) any weakly vanishing quantity
is strongly equal to a linear combination of constraints. Therefore, if
the quantity $R(q,p)$ is FC, it satisfies the strong equality
$\{ R,\vphi_s\} =R_s{^r}\vphi_r$. From this one can infer, by virtue of
the Jacoby identity, that the Poisson bracket of two FC constraints is
also FC.

The quantities $H'$ and $\phi_a$ in $H_T$ are FC. Thus, the number of
independent functions of time $v^a(t)$ in $H_T$ is equal to the number
of independent primary FC constraints $\phi_a$.

The presence of arbitrary multipliers in the equations of motion (and
their solutions) means that the variables $(q(t),p(t))$ can not be
uniquely determined from given initial values $(q(0),p(0))$; therefore,
they do not have a direct physical meaning. Physical informations about
a system can be obtained from functions $A(q,p)$, defined on constraint
surface, that are independent of arbitrary multipliers; such functions
are called (classical) {\it observables\/}. The physical state of a
system at time $t$ is determined by the complete set of observables at
that time.

In order to illustrate these ideas, let us consider a general dynamical
variable $g(t)$ at $t=0$, and its change after a short time interval
$\d t$. The initial value $g(0)$ is determined by $(q(0),p(0))$. The
value of $g(t)$ at time $\d t$ can be calculated from the equations of
motion: $g(\d t)=g(0)+\d t\{ g,H_T\}$. Since the coefficients $v^a(t)$
in $H_T$ are completely arbitrary, we can take different values for
these coefficients and obtain different values for $g(\d t)$, the
difference being of the form $\D g(\d t)=\ve^a\{ g,\phi_a\}$, where
$\ve^a=\d t (v_2^a-v_1^a)$. This change of $g(\d t)$ is unphysical, as
$g_1(\d t)$ and $g_2(\d t)=g_1(\d t)+\D g(\d t)$ correspond to the same
physical state. We come to the conclusion that primary FC constraints
generate unphysical transformations of dynamical variables, known as
{\it gauge transformations\/}, that do not change the physical state of
the system.

Similarly, one can conclude that the quantity $\{\phi_a,\phi_b\}$ is
also the generator of unphysical transformations. Since $\phi_a$'s are
FC constraints,  their PB is strongly equal to a linear combinations of
FC constraints. We expect that this linear combination contains also
secondary FC constraints, and this is really seen to be the case in
practice. Therefore, secondary FC constraints are also the generators
of unphysical transformations.

These considerations do not allow us to conclude that {\it all\/}
secondary FC constraints generate unphysical transformations. Dirac
believed that this is true, but was unable to prove it (``Dirac's
conjecture"). The final answer to this long standing problem has been
given by Castellani \cite{12}.

The Hamiltonian dynamics based on $H_T$ is known to be
{\it equivalent\/} to the related Lagrangian dynamics.

\subsub{Extended Hamiltonian.} We have seen that gauge transformations,
generated by FC constraints, do not change the physical state of a
system. This suggests the possibility of generalizing the equations of
motion by allowing {\it any\/} evolution of dynamical variables that
does not change physical  states. To realize this idea one can
introduce the extended Hamiltonian,
\be
H_E=H' +v^{a}\phi_{a}+\l^{b}\chi_{b} \, ,                   \lab{3.8}
\ee
containing both primary ($\phi_{a}$) and secondary ($\chi_{b}$) FC
constraints. Here, all the gauge freedom is manifestly present in the
dynamics, and any difference between primary and secondary FC
constraints is completely absent.

The equations of motion following from $H_E$ are {\it not equivalent\/}
with the Euler--Lagrange equations, but the difference is unphysical.

\subsub{Dirac brackets.} The presence of second class constraints in
the theory means that there are dynamical degrees of freedom that are
of no importance. In order to be able to eliminate such variables it is
necessary to set up a new PB referring only to dynamically important
degrees of freedom.

After finding all FC constraints $\vphi_a$ $(a=1,2,...,N_1)$, the
remaining constraints $\th_s$ $(s=1,2,...,N_2)$ are second class, and
the matrix $\D_{rs}=\{\th_r,\th_s \}$ is nonsingular. The new PB is
called the Dirac bracket,
\be
\{ f,g\}^*=\{ f,g\}-\{ f,\th_r\}\D^{-1}_{rs}\{\th_s, g\}\, ,\lab{3.9}
\ee
and it satisfies all the standard properties of a PB.

The Dirac bracket of any second class constraint with an arbitrary
variable vanishes by construction. This means that after the Dirac
brackets are constructed, second class constraints can be treated as
strong equations. The equations of motion (3.5) can be written in terms
of the Dirac brackets as $\dot g\approx\{ g,H_T\}^*$.

\subsub{Gauge conditions.} Gauge symmetries describe unphysical
transformations of dynamical variables. This fact can be used to impose
suitable restrictions on the set of dynamical variables, so as to bring
them into a 1--1 correspondence with the set of all observables. By
means of that procedure one can remove unobservable gauge freedom in
the description of dynamical variables, without changing any observable
property of the theory. The restrictions are realized as a suitable set
of gauge conditions:
\be
\O_a (q,p)\approx 0       \qquad (a=1,2,...,N_{gc})\, .    \lab{3.10}
\ee
The number of gauge conditions must be equal to the number of
independent gauge transformations. In the extended Hamiltonian
formalism $N_{gc}=N_1$.

First and second class constraints, together with gauge conditions,
define a subspace $\G^*$ of the phase space $\G$, having dimension
$N^*=2N - (2N_1 + N_2)$, in which the dynamics of independent degrees
of freedom is realized. This counting of independent degrees of freedom
is gauge independent, hence it holds also in the total Hamiltonian
formalism.

\subsub{Generators of gauge symmetries.} The presence of arbitrary
multipliers in $H_T$ is a signal of the existence of gauge symmetries
in the theory. The general method for constructing the generators of
such symmetries has been developed by Castellani \cite{12}.

We are considering a theory determined by the total Hamiltonian
\eq{3.7} and a complete set of constraints $\vphi_s\approx 0$. Suppose
that we have a trajectory $T_1(t)=\left(q_1(t),p_1(t)\right)$, that
starts from a point $T_0=(q(0),p(0))$ in $\G_2$, and satisfies the
equations of motion with some fixed functions $v^a_1(t)$. Consider now
a new trajectory $T_2(t)=\left(q_2(t),p_2(t)\right)$, that starts from
the same point $T_0$, and satisfies the equations of motion with new
functions $v^a_2(t)$. Transition from one to the other trajectory at
the same moment of time represents an unphysical or gauge
transformation (Figure 2).
\begin{figure}
\begin{center}
\includegraphics[height=4cm]{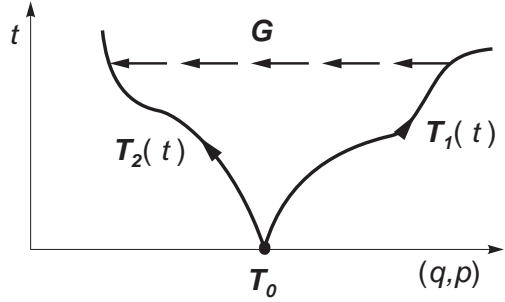}
\end{center}
\vspace*{-.5cm}
\caption{The Hamiltonian description of gauge transformations}
\end{figure}
If gauge transformations are given in terms of arbitrary parameters
$\ve(t)$ and their first time derivatives $\dot\ve(t)$, as is the case
with Poincar\'e gauge symmetry, the gauge generators have the form
\bsubeq\lab{3.11}
\be
G= \ve(t)G_0 + {\dot\ve}(t)G_1 \, ,                       \lab{3.11a}
\ee
where phase space functions $G_0$ and $G_1$ satisfy the conditions
\bea
             G_1&&=C_{PFC}\, ,\nn\\
G_0+\{ G_1,H_T\}&&=C_{PFC}\, ,\nn\\
    \{ G_0,H_T\}&&=C_{PFC}\, ,                            \lab{3.11b}
\eea
\esubeq
and $C_{PFC}$ denotes primary FC constraint. These conditions
clearly define the procedure for constructing the generator: one starts
with an arbitrary primary FC constraint $G_1$, evaluates its PB with
$H_T$, and defines $G_0$ in accordance with $\{ G_0,H_T\}=C_{PFC}$.

A detailed analysis, based on Castellani's algorithm, shows that in all
relevant physical applications Dirac's conjecture remains true. This is
of particular importance for the standard quantization methods, which
are based on the assumption that all FC constraints are gauge
generators.

\subsub{Electrodynamics.} Field theory can be thought of as a
mechanical system in which dynamical variables are defined at each
point $\mb{x}$ of a three--dimensional space,
$(q_i,p_i)\to(q_{ix},p_{ix})$ $\equiv (q_i(\mb{x}),p_i(\mb{x}))$, i.e.
where each index takes on also continuous values, $i\equiv (i,\mb{x})$.
Then, a formal generalization of the previous analysis to the case of
field theory becomes rather direct.

We now discuss the simple but important example of electrodynamics,
which is in many aspects similar to the theory of gravity. Dynamics of
the free electromagnetic field $A^\m(x)$ is described by the Lagrangian
$$
L= -\fr{1}{4} \int d^3x F_{\m\n}F^{\m\n}\, ,\qquad
    F_{\m\n}=\pd_\m A_\n-\pd_\n A_\m \, .
$$
Varying this Lagrangian with respect to $\dot A^\m$ one obtains the
momenta $\p_\m (x)= -F_{0\m}(x)$, where $x\equiv (t,\mb{x})$. Since
$F_{\m\n}$ is defined as the antisymmetric derivative of $A_\m$, it
does not depend on the velocity $\dot A^0$, so that the related
momentum vanishes. Thus, one obtains the primary constraint
$\vphi_1 \equiv \p_0 \approx 0$, while $\p_\a=-F_{0\a}$. With one
primary constraint present, the total Hamiltonian takes the form
$$
H_T=H_c+\int d^3x u(x)\p_0(x)\, ,
$$
where $H_c =\int d^3x\bigl(\fr{1}{4}F_{\a\b}F^{\a\b}
                          -\fr{1}{2}\p_\a\p^\a-A^0\pd^\a\p_\a\bigr)$.

By use of the basic PBs,
$\{A^\m(x),\p_\n(x')\}=\d^\m_\n\d(\mb{x}-\mb{x}')$, the consistency
condition for $\vphi_1$ leads to a secondary constraint:
$\vphi_2\equiv \pd^\a\p_\a \approx 0$. Further consistency requirement
on $\vphi_2$ is automatically satisfied. The constraints $\vphi_1$ and
$\vphi_2$ are FC, since $\{\p_0,\pd^\a\p_\a\}=0$.

The variable $A^0$ appears linearly in $H_c$. Its equation of motion,
$\dot A{^0}=\{A^0,H_T\}=u$, implies that $A^0$ is also an arbitrary
function of time. We observe that the secondary FC constraint $\vphi_2$
is already present in $H_c$ in the form $A^0\vphi_2$. In this way here,
as well as in gravitation, we find an interesting situation that all FC
constraints are present in $H_T$.

Let us now look for the generator of gauge symmetries in the form
\eq{3.11}. Starting with $G_1=\p_0$ one obtains
$$
G=\int d^3x (\dot\ve\p_0 -\ve\pd^\a\p_\a)\, ,
$$
so that the related gauge transformations are $\d_0 A^\m =\pd^\m \ve$,
$\d_0 \p_\m =0$. The result has the form we know from the Lagrangian
analysis.

The previous treatment fully respects the gauge symmetry of the theory.
This symmetry can be fixed by choosing two gauge conditions. The
dimension of the phase space $\G(A,\p)$ is eight; after fixing two
gauge conditions we come to the phase space $\G^*$ of dimension
$N^*=8-2\times 2 = 4$, corresponding to two Lagrangian degrees of
freedom of the massless photon.

\sect{Hamiltonian dynamics of PGT}

PGT represents a natural extension of the gauge principle to spacetime
symmetries. Now we want to present basic features of the Hamiltonian
approach to the general PGT. It leads to a simple form of the
gravitational Hamiltonian, representing a generalization of the
canonical Arnowitt--Deser--Misner (ADM) Hamiltonian from GR \cite{13},
and enables a clear understanding of the interrelation between
dynamical and geometric aspects of the theory \cite{14,15,16}.

\subsub{Primary constraints.} The geometric framework for PGT is
defined by the Riemann--Cartan spacetime $U_4$, while the general
Lagrangian has the form \eq{2.4}. The canonical momenta
$(\pi_k{^\m},\pi_{ij}{^\m},\pi)$, corresponding to the basic
Lagrangian variables $(b^k{_\m},A^{ij}{_\m},\psi)$, are obtained
from $\wtilde\cL$ in the usual manner.

Due to the fact that the curvature and the torsion are defined
through the antisymmetric derivatives of $b^k{_\m}$ and $A^{ij}{_\m}$,
respectively, they do not involve velocities of $b^k{_0}$ and
$A^{ij}{_0}$. As a consequence, one immediately obtains the following
set of the so--called {\it sure\/} primary constraints:
\be
\phi_k{^0}\equiv\pi_k{^0}\approx 0\, ,\qquad
   \phi_{ij}{^0}\equiv\pi_{ij}{^0}\approx 0 \, .            \lab{4.1}
\ee
These constraints are always present, independently of the values of
parameters in the Lagrangian. They are particularly important for
the structure of the theory. Depending on a specific form of the
Lagrangian, one may also have {\it additional\/} primary constraints.

\subsub{Hamiltonian.} The canonical Hamiltonian has the form
$\cH_c=\cH_M+\cH_G$, where $\cH_M=\pi\pd_0\psi-\wtilde\cL_M$,
and $\cH_G=\pi_k{^\a}\pd_0 b^k{_\a}
           +\fr{1}{2}\pi_{ij}{^\a}\pd_0 A^{ij}{_\a}-\wtilde\cL_G$.
The total Hamiltonian is defined by the expression
\be
\cH_T= \cH_c +u^k{_0}\phi_k{^0}
      +\fr{1}{2}u^{ij}{_0}\phi_{ij}{^0} +(u\cdot\phi)\, ,   \lab{4.2}
\ee
where $\phi$ denotes all additional primary constraints if they exist
(if--constraints).

The evaluation of the consistency conditions of the sure primary
constraints (4.1) is essentially simplified if we previously find out
the dependence of the Hamiltonian on the unphysical variables $b^k{_0}$
and $A^{ij}{_0}$. The analysis of this problem shows that $\cH_c$ is
linear in $b^k{_0}$ and $A^{ij}{_0}$, up to a three--divergence,
\bsubeq\lab{4.3}
\be
\cH_c=b^k{_0}\cH_k-\fr{1}{2}A^{ij}{_0}\cH_{ij}+\pd_\a D^\a\, ,
                                                           \lab{4.3a}
\ee
while possible extra primary constraints $\phi$ are independent of
$b^k{_0}$ and $A^{ij}{_0}$.

The above result may be put into a more geometric form. If \mb{n} is
the unit normal to the hypersurface $\S_0: x^0=\hbox{const.}$, the four
tangent vectors $(\mb{n},\mb{e}_\a)$ define the so--called ADM basis.
Any tangent vector $V_k$ can be expressed in terms of its orthogonal
and parallel components: $V_k=n_kV_\bot+V_{\bk}$, where
$V_\bot=n^kV_k$, $V_{\bk}=V_k-n_kV_\bot$, and $n^kV_{\bk}=0$. With
this notation, the above expression for $\cH_c$ can be rewritten in
an equivalent form:
\be
\cH_c =N\cH_\bot+N^\a\cH_\a
       -\fr{1}{2}A^{ij}{_0}\cH_{ij} + \pd_\a D^\a \, ,     \lab{4.3b}
\ee
\esubeq
where $N$ and $N^\a$ are lapse and shift functions, which are linear
in $b^k{_0}$: $N=n_k b^k{_0}$, $N^\a=h_{\bk}{^\a}b^k{_0}$.

The proof of this relation for the {\it matter\/} Hamiltonian goes as
follows. First, we decompose $\nabla_k\psi$ into its orthogonal and
parallel components, $\nabla_k\psi=n_k\nabla_\bot\psi+\nabla_{\bk}\psi$,
where complete dependence on velocities and unphysical variables
$(b^k{_0},A^{ij}{_0})$ is contained in $\nabla_\bot\psi$. Replacing
this into $\cL_M$ yields
$\cL_M=\bcL_M(\psi,\nabla_{\bk}\psi,\nabla_\bot\psi;n^k)$. Second,
using the factorization of the determinant, $b=\det (b^k{_\m})=NJ$,
where $J$ does not depend on $b^k{_0}$, the expression for $\pi$ can be
written in the form
$$
\pi={\pd\cL_M\over\pd\nabla_0\psi}
   =J{\pd\tilde\cL_M\over\pd\nabla_\bot\psi}\, .
$$
Finally, using the relation
$\nabla_0\psi=N\nabla_\bot\psi+N^\a\nabla_\a\psi$ to express the
velocity $\pd_0\psi$, the canonical Hamiltonian for matter fields takes
the form \eq{4.3b}, where
\bea
&&\cH_\a^M=\pi\nabla_\a\psi\, ,\qquad\cH_{ij}^M=\pi\S_{ij}\psi\, ,\nn\\
&&\cH_\bot^M=\pi\nabla_\ort\psi -J\bcL_M\, ,
               \qquad  D_M^\a = 0 \, .                       \lab{4.4}
\eea

Expressions for $\cH^M_\a$ and $\cH^M_{ij}$ are independent of
unphysical variables. They do not depend on the specific form of the
Lagrangian, but only on the transformation properties of the
fields, and are called {\it kinematical\/} terms of the Hamiltonian.
The term $\cH_\ort^M$ is {\it dynamical\/}, as it depends on the choice
of $\cL_M$. It represents the Legendre transformation of $\cL_M$ with
respect to the ``velocity" $\nabla_\ort\psi$. After eliminating
$\nabla_\ort\psi$ with the help of the relation defining $\pi$, one
finds that the dynamical Hamiltonian does not depend on unphysical
variables: $\cH_\ort^M=\cH_\ort^M(\psi,\nabla_{\bk}\psi,\pi /J;n^k)$.

If the matter Lagrangian is singular, the equations for momenta give
rise to additional primary constraints, which are again independent of
unphysical variables.

Construction of the {\it gravitational\/} Hamiltonian can be performed
in a very similar way, the role of $\nabla_k\psi$ being taken over by
$T^i{_{km}}$ and $R^{ij}{_{km}}$. In the first step we decompose
the torsion and the curvature, in last two indices, in the orthogonal
and parallel components. The parallel components $T^i{_{\bk\bm}}$ and
$R^{ij}{_{\bk\bm}}$ are independent of velocities and unphysical
variables. The replacement in the gravitational Lagrangian yields
$\cL_G=\bcL_G(T^i{_{\bk\bm}},R^{ij}{_{\bk\bm}},
               T^i{_{\bot\bk}},R^{ij}{_{\bot\bk}};n^k)$.
Using $b=NJ$ we find
$$
\hp_i{^{\bk}}=J{\pd\bcL_G\over\pd T^i{_{\ort\bk}}}\, ,\qquad
\hp_{ij}{^{\bk}}=J{\pd\bcL_G\over\pd R^{ij}{_{\ort\bk}}}\, ,
$$
where $\hp_i{^{\bk}}\equiv\pi_i{^\a}b^k{_\a}$ and
$\hp_{ij}{^{\bk}}\equiv\pi_{ij}{^\a}b^k{_\a}$  are ``parallel"
gravitational momenta. The velocities $\pd_0 b^i{_\a}$ and
$\pd_0 A^{ij}{_\a}$ can be calculated from the definitions of
$T^i{_{0\a}}$ and $R^{ij}{_{0\a}}$. After a straightforward algebra
the canonical Hamiltonian takes the form \eq{4.3b}, where
\bea
&&\cH_{ij}^G=2\pi_{[i}{^\a}b_{j]\a}+\nabla_\a\pi_{ij}{^\a}\, ,\nn\\
&&\cH_\a^G=\pi_i{^\b}T^i{_{\a\b}}+\fr{1}{2}\pi_{ij}{^\b}R^{ij}{_{\a\b}}
           -b^k{_{\a}}\nabla_\b\pi_k{^\b}\, ,\nn\\
&&\cH_\ort^G=\bigl( \hp_i{^{\bm}}T^i{_{\ort\bm}}
           +\fr{1}{2}\hp_{ij}{^{\bm}}R^{ij}{_{\ort\bm}}
           -J\bcL_G\bigr) -n^k\nabla_\b\pi_k{^\b}\, ,\nn\\
&&D_G^\a=b^i{_0}\pi_i{^\a}+\fr{1}{2}A^{ij}{_0}\pi_{ij}{^\a}\, .
                                                             \lab{4.5}
\eea
The expressions $T^i{_{\ort\bm}}$ and $R^{ij}{_{\ort\bm}}$
in $\cH_\ort^G$ should be eliminated with the help of the equations
defining $\hp_i{^{\bm}}$ and $\hp_{ij}{^{\bm}}$.

\subsub{Secondary constraints.} The basic result of the preceding
exposition is the conclusion that the canonical Hamiltonian has the
form \eq{4.3b}. Consequently, the consistency conditions of the sure
primary constraints  imply the following secondary constraints:
\be
\cH_\bot\approx 0\, ,\qquad
\cH_\a\approx 0\, ,  \qquad\cH_{ij}\approx 0\, .            \lab{4.6}
\ee

By working out the constraint algebra we shall see that the consistency
conditions of these constraints are automatically satisfied.

\subsub{Constraint algebra.} When no extra constraints are present in
the theory, the PB algebra of the secondary constraints \eq{4.6} takes
the form \cite{17}
\bsubeq\lab{4.7}
\bea
&&\{\cH_{ij},\cH'_{kl}\} =
  (\eta_{ik}\cH_{lj}-\eta_{jk}\cH_{li})\d-(k \lra l)\, ,\nn\\
&&\{\cH_{ij},\cH'_\a \} =0\, ,\nn\\
&&\{\cH_\a ,\cH'_\b \} =\bigl(\cH_\a '\pd_\b +\cH_\b\pd_\a
  -\fr{1}{2}R^{ij}{_{\a\b}}\cH_{ij}\bigr)\d\, ,           \lab{4.7a}
\eea
\bea
&&\{\cH_{ij},\cH'_\bot \} =0\, ,\nn\\
&&\{\cH_\a ,\cH'_\bot \} =\bigl(\cH_\bot\pd_\a
       -\fr{1}{2}R^{ij}{_{\a\bot}}\cH_{ij}\bigr)\d \, ,\nn\\
&&\{\cH_\bot ,\cH'_\bot \} =-\bigl({^3}g^{\a\b}\cH_\a +
       {^3}g^{\prime\a\b}\cH_\a '\bigr)\pd_\b\d\, .        \lab{4.7b}
\eea
\esubeq

The first three relations represent PBs between kinematical constraints
$\cH_{ij}$ and $\cH_\a$, the form of which does not depend of the
choice of the action. The relations \eq{4.7b} contain the dynamical part
of the Hamiltonian $\cH_\bot$.

Equations \eq{4.7} in the local Lorentz basis take the form
\bea
&&\{\cH_{ij},\cH_{kl}'\}=
      (\eta_{ik}\cH_{lj}-\eta_{jk}\cH_{li})\d-(k\lra l)\, ,\nn\\
&&\{\cH_{ij},\cH_k '\} =-(\eta_{ki}\cH_{j}-\eta_{kj}\cH_i)\d\, ,\nn\\
&&\{\cH_k,\cH_m '\} =-\bigl(n_{[m}R^{ij}{_{\bar k]\bot}}
    + \fr{1}{2} R^{ij}{_{\bar k\bar m}}\bigr)\cH_{ij}\d
    + 2\bigl(n_{[m}T^i{_{\bar k]\bot}}
    + \fr{1}{2}T^i{_{\bar k\bar m}}\bigr) \cH_i\d \, ,       \lab{4.8}
\eea
featuring a visible analogy with the standard Poincar\'e algebra.

When extra constraints are present the whole analysis becomes much more
involved, but the result essentially coincides with \eq{4.7}:
\bitem
\item[\bull] dynamical Hamiltonian $\cH_\bot$ goes over into a
redefined expression $\bcH_\bot$, that includes the contributions of
all primary second class constraints;
\item[\bull] the algebra may contain terms of the type $C_{PFC}$.
\eitem
Thus, secondary constraints \eq{4.6} are FC, and their consistency
conditions are automatically satisfied.

\sect{Gauge generators}

The generators of the Poincar\'e gauge symmetry have the form
$G=\dot\ve (t)G^{(1)}+\ve (t)G^{(0)}$, where $G^{(0)}, G^{(1)}$ are
phase space functions satisfying the conditions \eq{3.11b}. It is clear
that the construction of the gauge generators demands the knowledge of
the algebra of constraints. Since the Poincar\'e gauge symmetry is
always present, independently of a specific form of the action, one
naturally expects that all essential features of the Poincar\'e gauge
generators can be obtained by considering the simple case of the theory
with no extra constraints. After that, the obtained result is easily
generalized \cite{18}.

When no extra constraints exist, the primary constraints $\pi_k{^0}$
and $\pi_{ij}{^0}$ are FC. Starting with $G_k^{(1)}=-\pi_k{^0}$,
$G_{ij}^{(1)}=-\pi_{ij}{^0}$, and the related parameters $\xi^k$,
$\ve^{ij}$, the conditions \eq{3.11b} yield the following expression
for the Poincar\'e gauge generator  :
\bsubeq\lab{5.1}
\be
G=-\int d^3x \bigl[ \dot\xi^\m\bigl( b^k{_\m}\pi_k{^0}
  +\fr{1}{2}A^{ij}{_\m}\pi_{ij}{^0} \bigr)+\xi^\m\cP_\m
  +\fr{1}{2}\dot\o^{ij}\pi_{ij}{^0}+\fr{1}{2}\o^{ij}S_{ij}\bigr]\, ,
                                                           \lab{5.1a}
\ee
where $\xi^\m =\xi^kh_k{^\m}$ and $\o^{ij}=\ve^{ij}+\xi^\n A^{ij}{_\n}$
are new parameters, and
\bea
&&\cP_\m =b^k{_\m}\cH_k-\fr{1}{2}A^{ij}{_\m}\cH_{ij}
  +b^k{_{0,\m}}\pi_k{^0}+\fr{1}{2}A^{ij}{_{0,\m}}\pi_{ij}{^0}\, ,\nn\\
&&S_{ij}=-\cH_{ij}+2b_{[i0}\pi_{j]}{^0}+2A^s{_{[i0}}\pi_{sj]}{^0} \, .
                                                           \lab{5.1b}
\eea
\esubeq
Note that the time translation generator is equal to the total
Hamiltonian, $\cP_0 =\what\cH_T \equiv \cH_T -\pd_\a D^\a$,
since $\dot b^k{_0}=u^k{_0}$, $\dot A^{ij}{_0}=u^{ij}{_0}$, on shell.

One can show that the action of the generator \eq{5.1} on the fields
$\psi, b^k{_\m}$ and $A^{ij}{_\m}$ produces the complete Poincar\'e
gauge transformations:
\bea
&&\d_0\psi=\fr{1}{2}\o^{ij}\S_{ij}\psi-\xi^\n\pd_\n\psi\, ,\nn\\
&&\d_0 b^k{_\m}=\o^k{_s}b^s{_\m}-\xi^\l{_{,\m}}b^k{_\l}
                  -\xi^\l\pd_\l b^k{_\m}\, ,\nn\\
&&\d_0 A^{ij}{_\m}=-\nabla_\m\o^{ij}-\xi^\l{_{,\m}}A^{ij}{_\l}
                  -\xi^\l\pd_\l A^{ij}{_\m} \, ,          \lab{5.2}
\eea
where $\d_0 Q\equiv\{ Q,G\}$. The only properties of the total
Hamiltonian used in the derivation are the following:
$a)$ $\cH_T$ does not depend on the derivatives of momentum
variables, and
$b)$ it governs the time evolution of dynamical variables, i.e.
$\dot Q=\{ Q,H_T\}$.

In a similar way one can verify that the generator \eq{5.1} produces
the correct transformations of momenta.

We note here that the field transformations \eq{5.2} are symmetry
transformations of the action not only in the simple case characterized
by the absence of any extra constraints, but also in the general case
when these constraints exist. This fact  leads to the conclusion that
the gauge generator \eq{5.1}, in which  the term $P_0$ is {\it
replaced\/} by the new $\what\cH_T$, is the correct gauge generator of
the Poincar\'e symmetry for {\it any\/} choice of parameters.

\sect{Conservation laws}

Physical content of the notion of symmetry depends not only on the
symmetry of the action, but also on the symmetry of boundary
conditions. The choice of the asymptotic behaviour of gravitational
variables defines the asymptotic structure of spacetime, i.e. the
vacuum. The symmetry of the action breaks down to the symmetry of the
vacuum (spontaneous symmetry breaking), which becomes the physical
symmetry and determines the conservation laws. Thus, the possibility of
defining the concept of energy (and other conserved quantities) depends
essentially on the structure of asymptotic symmetries.

Assuming that the asymptotic symmetry in PGT is the global
Poincar\'e symmetry, we discuss the form of the related generators.
A careful analysis of boundary conditions leads to the
appearance of certain {\it surface terms\/} in the expressions for the
generators. The improved generators enable a correct treatment of the
conservation laws of energy, momentum and angular momentum \cite{19,20}.

\subsub{Asymptotic symmetry.} The global Poincar\'e transformations of
fields in the asymptotic region can be obtained from the corresponding
gauge transformations by the following replacement of parameters:
$$
\o^{ij}(x)\to \o^{ij}\, ,\qquad
\xi^\m (x)\to \o^\m{_\n}x^\n+\ve^\n\equiv\xi^\m\, ,
$$
where $\o^{ij}$ and $\ve^\n$ are constants. The generator of these
transformations can be obtained from the gauge generator \eq{5.1} in
the same manner, leading to
\be
G= \fr{1}{2}\o^{ij}M_{ij}-\ve^\n P_\n \, ,                  \lab{6.1}
\ee
where
\bea
&&P_{\m}=\int d^3x \cP_\m ,\qquad
 M_{\a\b}=\int d^3x\bigl(x_\a\cP_\b-x_\b\cP_\a-S_{\a\b}\bigr)\, ,\nn\\
&&M_{0\b}=\int d^3x\bigl(x_0\cP_\b-x_\b\cP_0-S_{0\b}+b^k{_\b}\pi_k{^0}
 +\fr{1}{2}A^{ij}{_\b}\pi_{ij}{^0}\bigr) \, .\nn
\eea

As the symmetry generators act on basic dynamical variables via PBs,
they are required to have {\it well defined functional derivatives\/}.
In case of parameters which decrease sufficiently fast at spatial
infinity, all partial integrations in $G$ are characterized by
vanishing surface terms, and the  differentiability of $G$ does not
represent any problem. The parameters of the global Poincar\'e
symmetries are not of that type, so that surface terms  must be treated
more carefully. We shall try to improve the form of the generators
\eq{6.1} so as to obtain the expressions with well defined functional
derivatives. The first step in that direction is to define precisely
the phase space in which the generators \eq{6.1} act.

\subsub{The phase space.} The choice of asymptotics becomes more clear
if we express the asymptotic structure of spacetime in certain
geometric terms. Here we shall consider isolated physical systems
characterized by matter fields which decrease sufficiently fast at
large distances, so that their contribution to surface integrals
vanishes. The spacetime outside an isolated system is said to be
{\it asymptotically flat\/} if the following two conditions are
satisfied:
\bitem
\item[\bull] $g_{\m\n}=\eta_{\m\n}+\cO_1$, where $\cO_n=\cO (r^{-n})$
denotes a term which decreases like $r^{-n}$ or faster for large $r$,
and $r^2=(x^1)^2+(x^2)^2+(x^3)^2$;
\item[\bull] $R^{ij}{_{\m\n}}=\cO_{2+\a}$ $(\a >0)$; this defines the
absolute parallelism in the asymptotic region.
\eitem

The second condition can be easily satisfied by demanding
$A^{ij}{_\m}=\cO_{1+\a}$. In the Einstein--Cartan (EC) theory, defined
by the simple action $\sim\int d^4x\,bR$, the connection behaves as the
derivative of the metric, i.e. $A=\cO_2$. We study here, for simplicity,
the EC theory, i.e. we assume that the {\it gravitational field\/}
behaves as
\be
b^k{_\m}=\d^k_\m +\cO_1 \, ,\qquad A^{ij}{_\m}=\cO_2 \, .   \lab{6.2}
\ee
In order to ensure the global Poincar\'e invariance of these conditions
we demand $b^k{_{\m ,\n}}=\cO_2$, $A^{ij}{_{\m ,\n}}=\cO_3$, etc.
The above requirements are minimal in the sense that some
additional arguments may lead to a better asymptotics, i.e. to a faster
or more precisely defined decrease of fields and their derivatives.

It should be noted that for those expressions that vanish on shell one
can demand an arbitrarily fast asymptotic decrease, as no solutions of
the field equations are thereby lost. In accordance with this, the
asymptotic behavior of {\it momenta\/} is determined by requiring
$p-{\pd\cL/\pd{\dot q}}=\what\cO$, where $\what\cO$ denotes a term that
decreases sufficiently fast. Using the definitions of the gravitational
momenta in EC theory one finds \cite{21}
\be
\pi_k{^0},\pi_{ij}{^0}=\what\cO \, ,\qquad
\pi_i{^\a}=\what\cO \, ,\qquad
\pi_{ij}{^\a}=-4aJn_{[i}h_{j]}{^\a}+\what\cO \, ,           \lab{6.3}
\ee
where $\k\equiv 1/2a$ is Einstein's gravitational constant.
Similar arguments yield the consistent asymptotic behavior of the
Hamiltonian multipliers.

\subsub{Improving the Poincar\'e generators.} The generators act on
dynamical variables via the PB operation,  which is defined in terms
of functional derivatives. A functional
$F[\vphi ,\pi ]=\int d^3x f\bigl(\vphi,\pd\vphi,\pi,\pd\pi\bigr)$
has well defined functional derivatives if its variation can be written
in the form $\d F=\int d^3x\bigl[ A(x)\d\vphi (x)+B(x)\d\pi (x)\bigr]$,
where terms $\d\varphi_{,\m}$ and $\d\pi_{,\m}$ are {\it absent\/}.
The global Poincar\'e generators do not satisfy this requirement.
However, this problematic behaviour can be simply corrected by adding
certain surface terms.

Let us see how this procedure works in the case of global
{\it spatial translatios\/}. The variation of $P_\a$ can be
written in the form
\bsubeq\lab{6.4}
\be
\d P_\a =-\d E_\a +R \, ,\qquad
E_\a\equiv\oint ds_\g\bigl(\pi_{ij}{^\b}A^{ij}{_{[\a}}\d_{\b ]}{^\g}
                                               \bigr) \, ,\lab{6.4a}
\ee
where $R$ denotes regular terms, not containing $\d\vphi_{,\m}$,
$\d\pi_{,\m}$, and the integration domain in $E_\a$ is the boundary of
the three--dimensional space. After this we can redefine $P_\a$,
\be
P_\a\to\wtilde P_\a\equiv P_\a +E_\a\, ,                  \lab{6.4b}
\ee
\esubeq
so that $\tilde P_\a$ has well defined functional derivative. One
can check that the assumed asymptotic behavior of phase--space
variables ensures finiteness of $E_\a$, which represents the
value of linear momentum.

The improved expression for the {\it time translation\/} generator takes
the form
\be
\wtilde P_0= P_0+E_0\, ,\qquad
E_0\equiv\oint ds_\g(-2aJh_a{^\a}h_b{^\g}A^{ab}{_\a} )\, . \lab{6.5}
\ee
The surface term $E_0$ is finite on account of the adopted asymptotics,
and it represents the value of the energy of the system.

In a similar way we can improve the form of the {\it spatial rotation\/}
and {\it boost\/} generators by introducing the following surface terms   :
\bea
&&E_{\a\b}\equiv\oint ds_\g \bigl[ -\pi_{\a\b}{^\g}
  +x_{[\a}\bigl( \pi_{ij}{^\g}A^{ij}{_{\b ]}}\bigr) \bigr]\, ,\nn\\
&&E_{0\b}\equiv\oint ds_\g \bigl[ -\pi_{0\b}{^\g}
  +x_0 \bigl(\pi_{ij}{^\a}A^{ij}{_{[\b}}\d_{\a ]}{^\g}\bigr)
  -x_\b\bigl( 2aJh_a{^\a}h_b{^\g}A^{ab}{_\a}\bigr)\bigr]\, . \lab{6.6}
\eea

A detailed analysis shows that the adopted asymptotic conditions do not
guarantee the finiteness of $E_{\a\b}$, as the integrand contains
$\cO_1$ terms. These troublesome terms are seen to vanish if we impose
the asymptotic gauge condition $a_{[ij]}=\cO_2$ on the gauge potentials
$a^k{_\m}=b^k{_\m}-\d^k_\m$, and  certain parity conditions. These
conditions are invariant under the global Poincar\'e transformations,
and they restrict the remaining gauge symmetry.  After that the surface
term $E_{\a\b}$ is seen to be finite, and, consequently,
$\wtilde M_{\a\b}$ is well defined.

All these results are obtained in the EC theory. Analogous
considerations in the general PGT show that the boost generator cannot
be redefined by adding a surface term. Therefore, it is not a well
defined generator under the adopted boundary conditions, but a
refinement of these could, in principle, lead to a correctly defined
boost generator.

\subsub{Conservation laws.} The improved generators satisfy the
standard PB algebra, up to squares (or higher powers) of constraints
and surface terms; this proves the asymptotic Poincar\'e invariance of
the theory. We now wish to see whether this symmetry implies, as usual,
the existence of certain conserved quantities.

After a slight modification Castellani's method can be applied to
study global symmetries, too. One can prove that necessary and
sufficient conditions that a phase--space functional $G(q,\pi,t)$
should be a generator of global symmetries take the form
\be
\{G,\wtilde H_T\} +{\pd G\over\pd t}= C_{PFC}\, ,\qquad
\{G,\vphi_s\}\approx 0 \, ,                                \lab{6.7}
\ee
where $\wtilde H_T$ is the improved Hamiltonian, $\vphi_s\approx 0$
are all constraints in the theory and, as before, the equality means
an equality up to the zero generators.

The improved Poincar\'e generators are easily seen to satisfy the
second condition, as they are given, up to surface terms, by volume
integrals of FC constraints. The first condition represents the
Hamiltonian form of the conservation law, since $dG/dt\equiv
\{G,\wtilde H_T\}+\pd G/\pd t$. It can be used to explicitly check the
conservation of $\wtilde P_\m$ and $\wtilde M_{\m\n}$.

Let us begin with the energy. One can easily see that
$d\wtilde P_0/dt=C_{PFC}$. Indeed, $\{\wtilde P_0,\wtilde P_0\} =0$,
and the only explicit time dependence is due to the presence of arbitrary
multipliers, $\pd\wtilde P_0/\pd t=\int d^3x\dot v^a\varphi_a=C_{PFC}$.
Hence,
$$
{d\wtilde P_0\over dt}\approx {d E_0\over dt}\approx 0 \, ,
$$
and we see that $E_0$ represents the value of the energy as a
conserved quantity.

The linear momentum $\wtilde P_\a$ and the spatial angular momentum
$\wtilde M_{\a\b}$ do not depend explicitly on time, and their
conservation is verified in a similar manner. The boost generator has
an explicit linear dependence on time, and
${d\wtilde M_{0\b}/dt}\approx {dE_{0\b}/dt}\approx -E_\b$. Hence, since
$E_\b\ne 0$ in general, the boost generator is not a conserved quantity
(it is conserved only in a reference frame where $E_\b =0$). This is a
consequence of an explicit linear time dependence of $\wtilde M_{0\b}$,
and the existence of a nonvanishing surface term in $\wtilde P_\b$.

\subsub{Comments.} In order to compare these results with those
obtained by the Lagrangian treatment, one should express all momentum
variables in $E_\m$ and $E_{\m\n}$ in terms of fields and their
derivatives, with the help of the constraints and the equations of
motion. A direct calculation shows that the energy--momentum and
angular momentum in EC theory coincide with the related GR expressions.

In the general $R+T^2+R^2$ theory one obtains the same results only
when all tordions ($A$--fields) are massive. The existence of massless
tordions has the following consequences: $a)$ the spatial angular
momentum $E_{\a\b}$ becomes different from the GR expression, and $b)$
the boost $E_{0\b}$ is not even defined in that case. The fact that the
boost generator is always defined in the Lagrangian approach can be
understood by observing that there one does not take care of the
existence of functional derivatives of the symmetry generators.

\sect{The teleparallel form of GR}

General geometric arena of PGT, the Riemann--Cartan space $U_4$, may be
a priori restricted by imposing certain conditions on the curvature and
the torsion.  An interesting limit of PGT is teleparallel or
Weitzenb\"ock geometry $T_4$, defined by the requirement
\be
R^{ij}{}_{\m\n}(A)=0 \, .                                  \lab{7.1}
\ee
The parallel transport in $T_4$ is path independent, i.e. we have
an absolute parallelism. The teleparallel geometry
is, in a sense, complementary to Riemannian geometry: while the
curvature vanishes, the connection may have torsion (Figure 1). Of
particular importance for the physical interpretation of this geometry
is the fact that there is a one--parameter family of teleparallel
Lagrangians which is empirically equivalent to GR \cite{5,22}. For the
parameter value $B=1/2$ the Lagrangian of the theory coincides, modulo
a four--divergence, with the Einstein--Hilbert Lagrangian, and defines
the teleparallel form of GR, \tgr.

The teleparallel description of gravity has been one of the most
promising alternatives to GR. However, analyzing this theory
Kopczy\'nsky \cite{23} found a hidden gauge symmetry, and concluded
that the torsion evolution is not completely determined by the field
equations. Assuming, then, that the torsion should be a measurable
physical quantity, he argued that this theory is internally
inconsistent (see also Refs. \cite{24} and \cite{25}). Here we focus our
attention on \tgr, showing how Dirac's canonical approach leads to a
clear explanation of this, somewhat mysterious behaviour \cite{26}.

\subsub{Lagrangian.} Gravitational dynamics in the framework of the
teleparallel geometry in PGT is described by a class of Lagrangians
quadratic in the torsion \cite{5,22}
\bea
&&\tcL = b\bigl(\cL_T +\cL_M\bigr)
          + \l_{ij}{}^{\m\n}R^{ij}{}_{\m\n}\, ,\nn\\
&&\cL_T=a\bigl(AT_{ijk}T^{ijk}+BT_{ijk}T^{jik}+CT_{k}T^{k}\bigr)
           \equiv \b_{ijk}(T)T^{ijk} \, ,                   \lab{7.2}
\eea
where the Lagrange multipliers $\l_{ij}{}^{\m\n}$ ensure the
teleparallelism condition \eq{7.1}, $a=1/2\k$, $T_k=T^m{}_{mk}$, and
$\cL_M$ is the Lagrangian of matter fields. Note that, here,
$\l_{ij}{}^{\m\n}$ is assumed  to be a tensor density rather then a
tensor, which simplifies the constraint analysis \cite{26}.

If we require that the theory \eq{7.2} describes all the standard
gravitational tests correctly, we can restrict our considerations to
the one--parameter family of Lagrangians, defined by the conditions
$(i)$ $\,2A+B+C=0\, ,C=-1$ \cite{5,22}.
This family represents a viable gravitational theory for macroscopic,
spinless matter, empirically indistinguishable from GR. Von der Heyde
\cite{27} and Hehl \cite{5} have given certain theoretical arguments in
favor of the choice $B=0$. There is, however, another, particularly
interesting choice determined by the requirement
$(ii)$ $\,2A-B=0$.
It leads effectively to the Einstein--Hilbert Lagrangian
$\cL_{GR}=-abR(\D)$, defined in Riemann spacetime $V_4$ with
Levi--Civit\`a connection $A=\D$, via the geometric identity:
$$
bR(A)= bR(\D)+b\bigl( \fr{1}{4}T_{ijk}T^{ijk}
      +\fr{1}{2}T_{ijk}T^{jik}-T_{k}T^{k}\bigr)-2\pd_\n(bT^{\n})\, .
$$
Indeed, in Weitzenb\"ock spacetime the above identity in conjunction
with the condition \eq{7.1} implies that the torsion Lagrangian
\eq{7.2} is equivalent to the Einstein--Hilbert Lagrangian, up to a
four--divergence, provided that
\be
A=\fr{1}{4}\, ,\qquad B=\fr{1}{2}\, ,\qquad C=-1  \, ,      \lab{7.3}
\ee
which coincides with the above conditions $(i)$ and $(ii)$.

The theory defined by Eqs. \eq{7.2} and \eq{7.3} is called the
teleparallel formulation of GR (\tgr). It is equivalent to GR for
scalar matter, but the gravitational couplings to spinning matter
fields in $T_4$ and $V_4$ are in general different. In what follows we
shall investigate the Hamiltonian structure and gauge properties of
\tgr\ without matter fields, but the results obtained here will be also
useful for the analysis of interacting theory.

\subsub{Primary constraints.} We begin the canonical analysis by
observing the presence of the primary constraints \eq{4.1}. Similarly,
the absence of the time derivative of $\l_{ij}{}^{\m\n}$ implies
\be
\phi^{ij}{}_{\m\n}\equiv \p^{ij}{}_{\m\n} \approx 0\, .      \lab{7.4}
\ee
The next set of constraints follows from the linearity of the curvature
in $\dot A^{ij}{}_\a$:
\be
\phi_{ij}{}^\a\equiv\p_{ij}{}^\a-4\l_{ij}{}^{0\a}\approx 0\, .\lab{7.5}
\ee

The system of equations defining $\hp_{i\bk}=\p_i{^\a}b_{k\a}$ can be
decomposed into irreducible parts with respect to the group of
three--dimensional rotations in $\S_0$. Taking into account the special
choice of parameters adopted in equation \eq{7.3}, we obtain two sets
of relations: the first set represents {\it extra\/} primary constraints
(if--constraints),
\bsubeq\lab{7.6}
\be
\hp_{\ort\bk}/J +2aT^\bm{}_{\bm\bk}\approx 0\, ,\qquad
\hp_{[\bi\bk]}/J-aT_{\ort\bi\bk}\approx 0\, ,               \lab{7.6a}
\ee
while the second set gives nonsingular equations, which can be
solved for velocities. Further calculations are greatly simplified by
observing that the set of extra constraints can be represented in a
unified manner as
\be
\phi_{ik}=\hp_{i\bk}-\hp_{k\bi}+a\nabla_\a B^{0\a}_{ik}\, ,\qquad
 B^{0\a}_{ik}\equiv\ve^{0\a\b\g}\ve_{ikmn}b^m_\b b^n_\g\, .\lab{7.6b}
\ee
\esubeq

\subsub{Hamiltonian.} Having found all the primary constraints, we can
now calculate the canonical Hamiltonian density. Following the usual
prescription \cite{26} we obtain
\bsubeq\lab{7.7}
\bea
\cH_c= N\cH_\ort+N^\a\cH_\a-\fr{1}{2}A^{ij}{_0}\cH_{ij}
      -\l_{ij}{}^{\a\b}R^{ij}{}_{\a\b} +\pd_\a D^\a \, ,  \lab{7.7a}
\eea
where
\bea
&&\cH_{ij}=2\pi_{[i}{^\a}b_{j]\a}+\nabla_\a\pi_{ij}{^\a} \, ,\nn\\
&&\cH_\a=\pi_i{^\b}T^i{_{\a\b}}-b^k{_\a}\nabla_\b\pi_k{^\b}\, ,\nn\\
&&\cH_\ort =\fr{1}{2}P_T^2-J\bcL_T(\bT)-n^k\nabla_\b\p_k{^\b}\, ,
                                                          \lab{7.7b}
\eea
$D^\a$ is the same as in Eq.\eq{4.5}, $\bT_{ijk}=T_{i\bj\bk}$, and
\bea
P_T^2\,&&={1\over 2aJ}\left( \hp_{(\bi\bk)}\hp^{(\bi\bk)}
        -{1\over 2}\hp^\bm{_\bm}\hp^\bn{_\bn} \right)\, ,\nn\\
\bcL_T(\bT)\,&&=a\left(\fr{1}{4}T_{m\bn\bk}T^{m\bn\bk}
        +\fr{1}{2}T_{\bm\bn\bk}T^{\bn\bm\bk}
        -T^\bm{}_{\bm\bk}T_\bn{}^{\bn\bk} \right) \, .     \lab{7.7c}
\eea \esubeq
Note minor changes in these expressions as compared to Eqs.\eq{4.3b}
and \eq{4.5}, which are caused by the presence of the Lagrange
multipliers in Eq.\eq{7.2}. The canonical Hamiltonian is now linear in
unphysical variables $b^k{_0}, A^{ij}{_0}$ {\it and\/}
$\l_{ij}{}^{\a\b}$.

The general Hamiltonian dynamics is described by the total Hamiltonian:
\be
\cH_T=\cH_c +u^i{_0}\p_i{^0}+\fr{1}{2}u^{ij}{_0}\p_{ij}{^0}
      +\fr{1}{4}u_{ij}{}^{\m\n}\p^{ij}{}_{\m\n}\nn\\
      +\fr{1}{2}u^{ik}\phi_{ik}
               +\fr{1}{2}u^{ij}{_\a}\phi_{ij}{^\a}\, .     \lab{7.8}
\ee

Although the torsion components $T_{\ort\ort\bk}$ and
$T^A_{\bi\ort\bk}$ are absent from the canonical Hamiltonian, they
reappear in the total Hamiltonian as nondynamical multipliers. Indeed,
the Hamiltonian field equations for $b^k{_\a}$ imply
$NT_{\ort\ort\bk}=u_{\ort\bk}$, $NT^A_{\bi\ort\bk}=u_{\bi\bk}$. Thus,
the existence of nondynamical torsion components is a phenomenon which
is equivalent to the presence of extra FC constraints $\phi_{ik}$.

\subsub{Consistency conditions.} Having found the form of the primary
constraints displayed in Eqs. \eq{4.1}, \eq{7.4}, \eq{7.5} and
\eq{7.6}, we now consider the requirements for their consistency.

The consistency conditions of the constraints \eq{4.1} have the form
\eq{4.6}. Similarly,
\bea
&&\dot\p^{ij}{}_{\a\b}\approx 0\quad\Ra\quad
     \chi^{ij}{}_{\a\b}\equiv R^{ij}{}_{\a\b}\approx 0\, ,\nn\\
&&\dot\p^{ij}{}_{0\b}\approx 0\quad\Ra\quad
         u^{ij}{}_{\b}\approx 0\, .                        \lab{7.9}
\eea
Since the equation of motion $\dot A^{ij}{_\b}=\{A^{ij}{_\a},\cH_T\}$
implies $R^{ij}{}_{0\b}\approx u^{ij}{_\b}$, all components of the
curvature tensor weakly vanish, as one could have expected.

The consistency condition for $\phi_{ij}{}^\a$ can be used to determine
$u_{ij}{}^{0\a}$:
\bea
4\bu_{ij}{}^{0\a}\approx &&\, N'\{\p_{ij}{^\a},\cH'_\ort\}
 -N^\a(\hp_{i\bj}-\hp_{j\bi})
 -A^{kl}{_0}(\eta_{ik}\p_{lj}{^\a}-\eta_{jk}\p_{li}{^\a})\nn\\
&& -4\nabla_\b\l_{ij}{}^{\b\a}
 +a(u_i{^s}B_{sj}^{0\a}-u_j{^s}B_{si}^{0\a})\, ,            \lab{7.10}
\eea
where a bar over $u$ is used to denote the determined multiplier.
Thus, the total Hamiltonian can be written in the form \eq{7.8} with
$u^{ij}{_\a}=0$ and $u_{ij}{}^{0\a}\to\bu_{ij}{}^{0\a}$.
Since $\bu_{ij}{}^{0\a}$ is linear in the multipliers
$(N,N^\a,A^{ij}{_0},\l_{ij}{}^{0\a},u_{kl})$, the total Hamiltonian
takes the form
\bsubeq\lab{7.11}
\bea
\cH_T=\,&&\hat\cH_T+\pd_\a D^\a
        -\fr{1}{2}\pd_\a(\l_{ij}{}^{\a\b}\p^{ij}{}_{0\b})\, ,\nn\\
\hat\cH_T\equiv\,&&N\bcH_\ort+N^\a\bcH_\a-\fr{1}{2}A^{ij}{_0}\bcH_{ij}
             -\l_{ij}{}^{\a\b}\bchi^{ij}{}_{\a\b} \nn\\
           &&+\, u^i{_0}\p_i{^0}+\fr{1}{2}u^{ij}{_0}\p_{ij}{^0}
             +\fr{1}{4}u_{ij}{}^{\a\b}\p^{ij}{}_{\a\b}
             +\fr{1}{2}u^{ij}\bphi_{ij} \, ,               \lab{7.11a}
\eea
where
\bea
&&\bcH_\ort=\cH_\ort
  +\fr{1}{8}\{\pi_{ij}{^\a},\cH'_\ort\}\pi^{ij}{}_{0\a}\, ,\nn\\
&&\bcH_\a=\cH_\a
  -\fr{1}{8}(\hp_{i\bj}-\hp_{j\bi})\pi^{ij}{}_{0\a}\, ,\nn\\
&&\bcH_{ij}=\cH_{ij}+\fr{1}{2}\pi_{[i}{^s}_{0\a}\pi_{j]s}{^\a}\, ,\nn\\
&&\bchi^{ij}{}_{\a\b}=R^{ij}{}_{\a\b}
  -\fr{1}{2}\nabla_{[\a}\pi^{ij}{}_{0\b]}\, ,\nn\\
&& \bphi_{kl}=\phi_{kl}-\fr{1}{4}a \bigl(
   \pi_k{^s}{}_{0\a}B_{sl}^{0\a}+\pi_l{^s}{}_{0\a}B_{ks}^{0\a}\bigr)\, .
                                                           \lab{7.11b}
\eea
\esubeq

The consistency conditions for the tetrad constraints $\phi_{ij}$,
technically the most complicated ones, are found to be automatically
fulfilled: $\dot\phi_{ij}=\{\phi_{ij},\cH_T\}\approx 0$.

Thus, the only secondary constraints are \eq{4.1} and
$R^{ij}{}_{\a\b}\approx 0$. Their consistency conditions are found to
be identically satisfied; hence there are {\it no tertiary
constraints\/}.

\subsub{Constraints and gauge symmetries.} In the previous analysis we
found that \tgr\ is characterized by the following set of constraints:
\bitem
\item[] primary:$\quad\p_i{^0},\p_{ij}{^0},\phi_{ij},
        \phi_{ij}{^\a},\p^{ij}{}_{\a\b},\p^{ij}{}_{0\b}\,$;\\
     secondary:$\quad\cH_\ort,\cH_\a,\cH_{ij},\chi^{ij}{}_{\a\b}$.
\eitem
The essential dynamical classification of these constraints reads as
follows:
\bitem
\item[] first class:$\quad\p_i{^0},\p_{ij}{^0},\p^{ij}{}_{\a\b},
 \bphi_{ij},\bcH_\ort,\bcH_\a,\bcH_{ij},\bchi^{ij}{}_{\a\b}\,$; \\
       second class:$\quad\phi_{ij}{^\a},\p^{ij}{}_{0\b}\,$.
\eitem
The constraints $\phi_{ij}{^\a}$ and $\p^{ij}{}_{0\b}$ are
second class since $\{\phi_{ij}{^\a},\p^{kl}{}_{0\b}\}\not\approx 0$.
They can be used as strong equalities to eliminate $\l_{ij}{}^{0\a}$
and $\p^{ij}{}_{0\b}$ from the theory and simplify the exposition.
The FC constraints are identified by observing that they appear
multiplied by arbitrary multipliers in the total Hamiltonian \eq{7.11a}.

The algebra of FC constraints plays an important role not only
in the construction of classical gauge generators, but also in
studying quantum properties of \tgr, such as the BRST structure.
These important subjects deserve further investigation.

The complete set of gauge generators for \tgr\ can be constructed
starting from the primary FC constraints $\p_i{^0}$, $\p_{ij}{^0}$,
$\p^{ij}{}_{\a\b}$ and $\bphi_{ij}$. One should note that $\p_i{^0}$,
$\p_{ij}{^0}$ and $\p^{ij}{}_{\a\b}$ are always present in the
teleparallel geometry \eq{7.2}, while $\bphi_{ij}$ are typical
if--constraints. The role of $\p_i{^0}$ and $\p_{ij}{^0}$ in the
Poincar\'e gauge symmetry is well known, but the meaning of gauge
symmetries generated by $\pi^{ij}{}_{\a\b}$ and $\bphi_{ij}$ is not
completely clear \cite{26}. The related gauge features of \tgr\ should
be further analyzed.

The existence of extra FC constraints $\bphi_{ij}$ may be interpreted
as a consequence of the fact that the velocities contained in
$T_{\ort\ort\bk}$ and $T^A_{\bi\ort\bk}$ appear at most linear in the
Lagrangian and, consequently, remain arbitrary functions of time. Heht
et al. \cite{28} concluded that the initial--value problem for \tgr\
becomes well defined if these undetermined velocities are simply gauged
away, ensuring the new kinetic Hessian matrix to be nondegenerate.
However, this conclusion should not be taken as a definitive one before
investigating possible nonlinear constraint effects \cite{29,25}.

We note that Maluf \cite{30} tried to analyze the canonical structure
of \tgr\ by imposing the time gauge at the Lagrangian level. However,
his arguments concerning the necessity of the time gauge are not
acceptable: it is clear that this gauge may be useful, but certainly
not essential. After fixing the time gauge, he found the Hamiltonian
and derived the related set of constraints (which is not complete), but
was unable to calculate the constraint algebra unless imposing another
gauge condition. Thus, his analysis of the gauge structure of \tgr\
remaines rather unclear.

\sect{Concluding remarks}

In this paper we presented basic features of the Hamiltonian
structure of PGT using Dirac's method for constrained dynamical
systems.

1) We first gave a short review of PGT and Dirac's method, by focussing
our attention on those aspects that are essential for the analysis of
the canonical structure of PGT.

2) The general form of the canonical PGT Hamiltonian is displayed in
Eqs.\eq{4.3}. This important result shows that $\cH_c$ is linear in
unphysical variables $b^k{_0}$ and $A^{ij}{_0}$, whereupon one simply
obtains the secondary constraints $\cH_\ort\approx 0,\cH_\a\approx 0$
and $\cH_{ij}\approx 0$.

3) The PB algebra of the Hamiltonian constraints $\cH_\ort,\cH_\a$ and
$\cH_{ij}$ is found to have the general form shown in Eq.\eq{4.7}. This
result is essential for studying the canonical consistency of PGT.

4) The PB algebra of constraints is used to construct the general form
of the generators of local Poincar\'e symmetry.

5) Assuming that spacetime behaves asymptotically as Minkowski space
$M_4$, one obtains the corresponding global Poincar\'e generators. The
surface terms related to the improved form of these generators are seen
to represent the related conserved charges --- energy, momentum and
angular momentum.

6) The constrained Hamiltonian analysis of noninteracting \tgr\ is
carried out without any gauge fixing, leading to the standard,
consistent canonical behaviour (up to the question of possible
nonlinear constraint effects). Extra gauge symmetries are found as a
specific feature of \tgr, and we are now investigating the conservation
laws \cite{31}.

\ack{Acknowledgments}

I wish to thank M. Vasili\'c for useful discussions.
This work was partially supported by the Serbian Science Foundation,
Yugoslavia.

\end{document}